# The 2/3 Rule of Glass Physics Implies Universalities in Crystal Melting

Peter Lunkenheimer,[1,*] Konrad Samwer,[2] and Alois Loidl[1]

[1] *Experimental Physics V, Center for Electronic Correlations and Magnetism, University of Augsburg, 86135 Augsburg, Germany*
[2] *1. Physikalisches Institut, University of Göttingen, 37077 Göttingen, Germany*

\* Contact author: peter.lunkenheimer@physik.uni-augsburg.de

Since more than 100 years, melting is thought to be governed by the Lindemann criterion. It assumes that a crystal melts when, upon heating, the growing atomic vibration amplitudes become sufficiently large to destabilize its crystalline lattice. However, it is unclear why the viscosities $\eta$ or the related relaxation times $\tau$ of the resulting liquids, measured directly at the melting point $T_m$, differ by up to nine decades, depending on the material. Based on the empirical rule that the ratio of the glass-transition temperature and $T_m$ is about 2/3, here we show that this strong variation is due to differences in the liquid's fragilities, a property associated with pronounced non-Arrhenius behavior and often ascribed to cooperative motions. We propose that, without cooperativity, all crystals would melt into liquids with a universal viscosity value and relaxation time. Hence, the real melting point is only partly determined by the Lindemann criterion and strongly enhanced by the cooperativity of the resulting liquid. Our findings are corroborated by the determination of the idealized, fragility-free melting temperatures, and of the corresponding $\eta$ and $\tau$ values for various example materials.

## I. INTRODUCTION

The viscosity $\eta$ quantifies the ability of a material to flow in response to an external force. It is approximately proportional to the relaxation time $\tau$, characterizing particle mobility. Remarkably, the absolute values of these quantities close to the melting point, $\eta_m := \eta(T_m)$ and $\tau_m := \tau(T_m)$, can vary by up to nine decades for different materials (see Table S1 of the Supplemental Material (SM) [1], including references [2-43]). This strong variation is puzzling. Why has, e.g., benzophenone after melting a viscosity of similar order as water [38,44], while melted $SiO_2$ is much more viscous than honey taken from a fridge [4,45]? Of course, for the covalently bound $SiO_2$, $T_m$ is much higher than for the van-der-Waals system benzophenone, mainly due to the different bonding strengths. However, the huge variance in their $\eta_m$ values (Table S1) means that a benzophenone crystal survives heating up to such high temperatures that the particles in the finally formed liquid move extremely fast [37] ($\tau_m \approx 1 \times 10^{-10}$ s; Table S1), while a $SiO_2$ crystal already melts when the particle motion in the resulting liquid is many decades slower. Understanding this "earlier" melting of materials like $SiO_2$ is also of technical relevance as it can be assumed to affect the temperatures necessary for preparing the common silicate glasses used for windows, bottles, etc. The strong variation of $\eta_m$ and $\tau_m$ is also surprising in light of the finding that in liquids, various crossover effects and transitions between regimes of different transport or relaxation behavior are known to occur at certain universal, material-independent values of $\eta$ or $\tau$ [46,47,48].

Recently, we reported [42] that thermal-expansion results on crystals are consistent with an enhancement of $T_m$ caused by cooperative particle motions, a property primarily known from the physics of glass-forming liquids [49,50,51,52] but also applicable to ordinary liquids that do not easily form glasses. We rationalized this finding by a reduction of entropy, leading to an increase of the free energy in the liquid state [42]. Could different degrees of cooperativity be responsible for the strong material-dependent variation of $\eta_m$ and $\tau_m$? In other words, would this variation vanish and a universal $\eta_m^{id}$ and $\tau_m^{id}$ be observed if cooperativity could be "switched off"? Then the materials would melt at an "idealized" melting temperature $T_m^{id}$, only determined by the Lindemann criterion and not affected by cooperativity. In systems like benzophenone (so-called "fragile" materials [4,53]), high cooperativity can be assumed to shift the actual $T_m$ to temperatures far above $T_m^{id}$. Therefore, $\eta_m$ and $\tau_m$ of such materials are significantly reduced, bearing in mind that $\eta$ and $\tau$ strongly decrease with increasing temperature. Without this effect, $\eta_m$ and $\tau_m$ would be much higher, possibly approaching the same values as in systems where cooperativity is not important (termed "strong") and, thus, $T_m^{id} \approx T_m$. To confirm this scenario, in the present work we derive $\eta_m^{id}$ and $\tau_m^{id}$ based on the 2/3 rule and on the relation of $T_m$ and $T_m^{id}$, recently proposed in Ref. [42]. We check the validity of this approach for various example materials.

Cooperativity causing fragile behavior is an essential ingredient in various theories of the glass transition, the crossover of a supercooled liquid into a non-crystalline solid [49,50,52]. We use this term as defined in the pioneering work by Adam and Gibbs [54] and also considered in various later works (e.g., 49,50,51,52,55]). There a glass-forming material is assumed to consist of "cooperatively rearranging regions" (CRRs), in which the molecules move in a correlated way. The temperature dependence of cooperativity is believed to lead to fragility, i.e., deviations of the viscosity of liquids from Arrhenius behavior, $\eta \propto \exp[E/(k_B T)]$ (where $E$ is an energy barrier), expected for simple thermal activation of the associated particle motions. Such deviations can occur for all types of liquids but are best observed in liquids that can be easily supercooled, because there a broad temperature range, extending significantly below $T_m$ is available. At very high temperatures, essentially single-particle motions dominate because there the interactions leading to cooperative behavior can be neglected, compared to the high thermal energy. However, with decreasing temperature increasingly larger effective energy barriers emerge due to continuously rising cooperativity, i.e., a growth of the CRRs mentioned above, thus explaining the often observed deviations from Arrhenius behavior [49,51,56]. The theoretical foundations of this scenario are treated in detail in the mentioned work by Adam and Gibbs [54] and, in more advanced form, within the framework of the random first-order transition theory, merging the Adam-Gibbs theory with other, more recent ones such as the mode-coupling theory [55,57].

Notably, there are also various alternative theoretical approaches that are able to explain the typical non-Arrhenius behavior of most glass-forming liquids, without invoking prominent contributions from cooperativity effects (e.g., refs. [13,58,59,60]). Each of these competing models assumes a specific physical mechanism that gives rise to fragile behavior. The present work does not intend to clarify the validity of these various approaches but to discuss the importance of the fragility for the observed strong variation of $\eta_m$ and $\tau_m$.

## II. FRAGILITY AND ITS INFLUENCE ON VISCOSITY AT $T_M$

The degree of non-Arrhenius behavior can be quantified by the fragility index $m$ [2]. Within an Angell plot [61], $\eta$ or $\tau$ versus the $T_g$-scaled inverse temperature (Fig. 1), $m$ is the slope close to $T_g$ (cf. dotted line for $m = 50$). The solid lines in Fig. 1 are calculated by the Vogel-Fulcher-Tammann (VFT) formula, written in its modified form as proposed by Angell [4]

$$y = y_0 \exp\left(\frac{DT_{VF}}{T - T_{VF}}\right) \qquad (1)$$

where $y$ stands for $\eta$ or $\tau$. $T_{VF}$ is the Vogel-Fulcher temperature, and $D$ denotes the strength parameter, characterizing the deviations from Arrhenius behavior. Equation (1) is often used for the empirical description of $\eta(T)$ or $\tau(T)$ curves in supercooled liquids [4,49,51] and was theoretically rationalized in several theoretical approaches [54,62]. Its divergence at $T_{VF}$ may support theories assuming a hidden phase transition underlying the glass transition [51,52,54,57]. While Eq. (1) is the most employed function for fitting such non-Arrhenius behavior, also numerous other empirical or theoretically founded functions exist to describe $\eta(T)$ or $\tau(T)$ data [13,58,59,60,63]. We want to point out that the principal conclusions of our work do not rely on the occurrence of such a divergence, and here the VFT formula can be regarded as a formal parameterization of the experimental data only. The lines in Fig. 1 are shown for both $\eta$ and $\tau$ (right and left ordinate, respectively) and for different fragility indices. The strong deviations from Arrhenius behavior (straight line in this representation) for high $m$ values become obvious. Making the common assumption of $\eta(T_g) = 10^{12}$ Pa s or $\tau(T_g) = 100$ s and using $10^{-14}$ s or $10^{-4}$ Pa s as typical pre-exponential factors [2,4], leads to $m = m_{min} = 16$ for bare Arrhenius behavior. (For the viscosity, instead often $\eta_0 \approx 10^{-5}$ Pa s is assumed [4,49,53], which results in $m_{min} \approx 17$ for pure Arrhenius behavior. This minor 6% difference, reflecting a weak temperature dependence in the $\tau/\eta$ ratio [2], does not affect the conclusions of the present work.) It should be noted that Fig. 1 assumes $\tau \propto \eta$. While this proportionality often is not exactly fulfilled [64], the deviations are rather small when compared to the huge temperature-dependent variations of $\tau$ and $\eta$, extending over many decades. An example for the very similar temperature dependences of both quantities for two molecular glass formers is provided in Ref. [24].

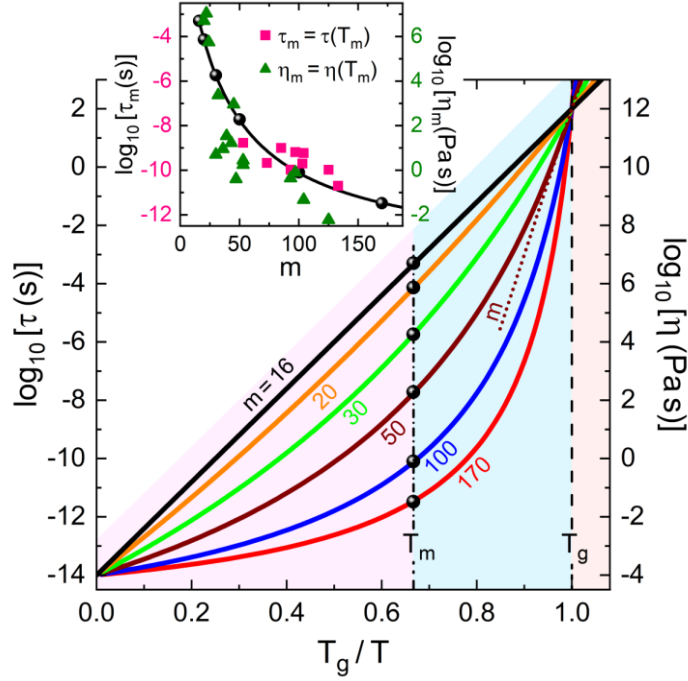

FIG. 1. Schematic Angell plot of $\tau$ (left ordinate) and $\eta$ (right ordinate) and their fragility dependence at $T_m$. The lines were calculated by Eq. (1) for different fragility values, indicated in the figure. For $m = 50$, the dotted line shows the slope at $T_g$ defining $m$ [2]. The dashed and dash-dotted lines indicate $T_g$ and $T_m$, respectively, obeying the 2/3 rule. The shaded areas visualize (from left to right) the temperature regions of the liquid, supercooled liquid, and glass states. The spheres represent the expected $\tau_m$ and $\eta_m$ values. The inset shows the $m$ dependence of $\tau_m$ and $\eta_m$ (left and right ordinate, respectively). The squares and triangles are experimental values (Table S1), the line was calculated using Eq. (3), and the spheres correspond to those in the main frame.

Figure 1 reveals a straightforward corroboration for the above-suggested influence of fragility on $\eta_m$ or $\tau_m$: According to the empirical "2/3 rule",

$$T_g \approx 2/3\, T_m, \qquad (2)$$

known in glass physics [11,22,42,49,65] [see Fig. 2(a) for a demonstration of its approximate validity], the melting point in Fig. 1 can be estimated by the vertical dash-dotted line [66]. As indicated by the spheres, then the absolute values of $\eta_m$ or $\tau_m$ should strongly depend on fragility. This remains valid, even when considering that Eq. (2) is often regarded as an approximate "rule of thumb" only and that the proportionality factor varies by about ±0.15, i.e., ~23 % [Fig. 2(b)]. To proceed further, we make use of the relation $D = m_{min}^2 \ln 10/(m - m_{min})$ from Ref. [2], relating the strength parameter of the VFT equation to the fragility index. It exactly follows from the VFT equation. Accounting for this relation and using Eqs. (1) and (2), $\tau_m(m)$ and $\eta_m(m)$ can be calculated (see SM [1]), leading to:

$$\log_{10} y_m = \frac{m_{min}^2}{m_{min} + m/2} + \log_{10} y_0 \qquad (3)$$

Here $y_m$ stands for $\tau_m$ or $\eta_m$ and $\log_{10} y_0$ is about -4 or -14 for $\eta$ or $\tau$, respectively [67]. This formula should be only approximately valid because the ratio $T_g/T_m$ varies by about 23 % [Fig. 2(b)], the preexponential factors can differ by 1-2 decades, and Eq. (1) does not always lead to perfect fits of the experimental data [23,68]. Nevertheless, as shown in the inset of Fig. 1, indeed the general trend of the experimental $\tau_m(m)$ and $\eta_m(m)$ data (symbols; Table S1) is reasonably matched by Eq. (3) (line). For systems with nearly pure Arrhenius behavior ($m = m_{min} \approx 16$), Eq. (3) predicts $\log_{10} \tau_m \approx -3.33$ or $\log_{10} \eta_m \approx 6.67$. Notably, the latter value is in accord with the experimental findings for the covalent-network glass formers $SiO_2$ and albite, which have very low fragilities of $m = 20$ and 22 and reveal experimental values of $\log_{10} \eta_m \approx 6.7$ and 7.0, respectively (Table S1).

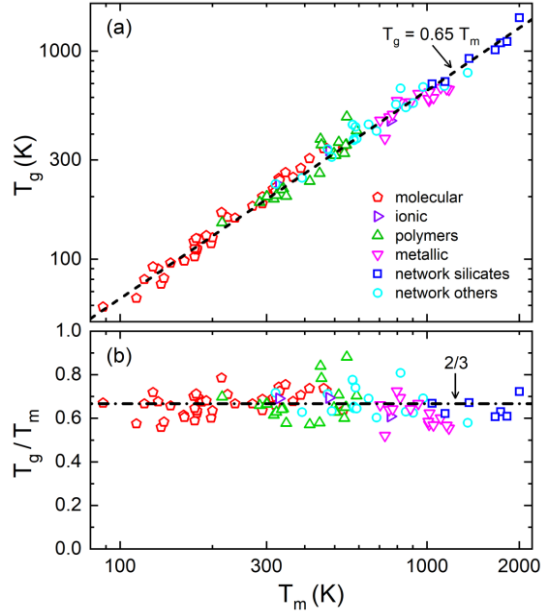

FIG. 2. Verification of the 2/3 rule. (a) $T_g$ versus $T_m$ plotted in a double-logarithmic representation for more than 80 materials from different material classes as listed in the legend [42] (see Supplemental Material of Ref. [42] for a list of all values and sources). The dashed line shows a fit with $T_g = a\, T_m$ (i.e., a straight line with slope one in this log-log plot), resulting in $a = 0.65$, close to the often-reported value $T_g/T_m \approx 2/3$. (b) Plot of $T_g/T_m$ versus $T_m$ (using a logarithmic scale for the latter), demonstrating the scatter of $a$. The dash-dotted line indicates the value 2/3.

## III. DERIVING UNIVERSALITIES FOR IDEALIZED MELTING

In Ref. [42], we found that the thermal expansion of crystals is not inversely proportional to $T_m$ but scales with $m$, analogous to earlier results for glasses and $T_g$ [69]. This implies that the Lindemann criterion, a well-established concept [70] assuming that crystals melt when the amplitudes of their atoms' thermal vibrations exceed a certain percentage of the lattice-site spacing [71,72], is insufficient to describe melting. Instead, melting is also strongly affected by the material's fragility. As shown in Ref. [42], the thermal-expansion data are consistent with an enhancement of $T_m$ by a factor $m/m_{\min}$, i.e.:

$$T_m = (m/m_{\min})\, T_m^{id}. \qquad (4)$$

The solid line in Fig. 3(a) shows an Arrhenius plot of $\tau(T)$ calculated from Eq. (1) using, as an example, approximate parameters as reported for glycerol [23] having intermediate fragility, $m \approx 53$. As discussed above, at high temperatures thermally activated single-particle motions should dominate, because there the cooperativity can be neglected, compared to the high thermal energy. Hence there, $y(T)$ should follow an Arrhenius law,

$$y = y_0 \exp\left(\frac{DT_{VF}}{T}\right), \qquad (5)$$

derived from Eq. (1) for $T \to \infty$. This leads to the hypothetical cooperativity-free $\tau(T)$ shown by the dashed straight line in Fig. 3(a). For alternative scenarios not considering cooperativity for the explanation of non-Arrhenius behavior, it also seems reasonable to assume Arrhenius temperature dependence at high temperatures. The particle motions determining viscosity are essentially thermally activated and, only upon cooling towards the glass temperature, deviations (no matter whether caused by cooperativity or not) develop. Indeed, high-temperature Arrhenius behavior was experimentally found for various materials [73,74,75] and used as a measure of fragility in Ref. [76].

As indicated in Fig. 3(a), after determining $T_g$ from $\tau(T_g) = 100$ s, one can apply Eq. (2) to deduce $T_m$ and then derive $T_m^{id}$ from Eq. (4). The star in Fig. 3(a) shows the resulting $\log_{10} \tau_m^{id} = -3.33$, the same value as deduced from Fig. 1 for $\tau_m$ of strong materials with $m = m_{\min}$. For the viscosity, similar considerations lead to $\log_{10} \eta_m^{id} = 6.67$, in accord with $\log_{10} \eta_m \approx 6.7$ experimentally found for $SiO_2$ whose $m \approx 20$ is close to $m_{\min}$ (Table S1) [4]. Consequently, we propose the existence of a universal, fragility-independent value of $\tau_m^{id}$ (and, correspondingly, $\eta_m^{id}$), a fundamental finding of the present work. Figure 3(b) shows the same construction for three different $m$ values within an Angell plot. Obviously, high fragilities imply large differences between $T_m$ (spheres) and $T_m^{id}$ (stars). Both temperatures approach each other for small fragilities, becoming identical for $m = m_{min}$. Identical values of $\log_{10} \tau_m^{id}$ (left ordinate) or $\log_{10} \eta_m^{id}$ (right ordinate) result for all fragilities (dash-dotted line).

The energy barriers $E$ (in K) of the dashed Arrhenius lines in Fig. 3 are equal to $DT_{VF}$ [Eq. (5)]. Notably, $E$ is proportional to $T_m^{id}$. This becomes obvious if one sets $T = T_m^{id}$ in Eq. (5), which leads to $\left(\log_{10} y_m^{id} - \log_{10} y_0\right) \ln 10 = E/T_m^{id}$, resulting in:

$$E = 24.6\, T_m^{id} \quad (6)$$

$E$ represents the energy barrier (in K) for fragility-free single-particle motion in the liquid. In accord with Eq. (6), it seems reasonable that $E$ is much higher than the thermal energy $k_B T_m^{id}$ needed to achieve 10-20% vibration amplitude in relation to the lattice-site spacing in the crystal, according to the Lindemann criterion [71,72].

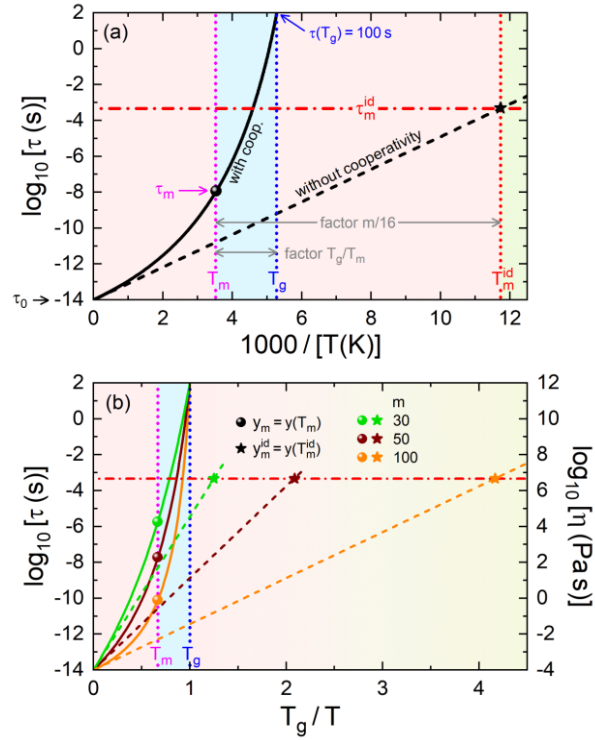

FIG. 3. (a) Arrhenius plot of $\tau(T)$. As an example, the solid line is calculated by Eq. (1) using $D = 15.8$ (implying $m = 53.3$) and $T_{VF} = 132$ K as reported for glycerol [23]. For $\tau_0$ the idealized value of $10^{-14}$ s is assumed. Dashed line: Arrhenius law, reflecting the hypothetical behavior without the effects leading to fragility [Eq. (5)]. The dotted lines show (from left to right) the characteristic temperatures $T_m$, $T_g$, and $T_m^{id}$. The dash-dotted line indicates $\tau_m^{id} = 10^{-3.33}$ s, deduced assuming $T_g/T_m = 2/3$ and $T_m = m/16\, T_m^{id}$. The sphere and star denote $\tau_m$ and $\tau_m^{id}$, respectively. (b) Angell plot of $\tau(T)$ (left ordinate) or $\eta(T)$ (right ordinate), calculated using Eq. (1) for three fragilities. The meaning of the lines and symbols are the same as in frame (a).

The above, essentially graphically based considerations can be confirmed by some straightforward calculations. We do this for free values of the preexponential factor $y_0$ and of $y(T_g)$ and define: $l_0 := \log_{10} y_0$ and $l_g := \log_{10}[y(T_g)]$ (cf. Fig. S1). Using rather simple algebra (see SM [1]), one then can derive:

$$\log_{10} y_m^{id} = \frac{T_g}{T_m}\left(l_g - l_0\right) + l_0 \quad (7)$$

Equation (7) implies that a fixed $T_g/T_m$ ratio leads to universal $\tau_m$ and $\eta_m$ values at the idealized melting temperature. Even when taking into account the mentioned scatter of $T_g/T_m$ of about ±0.15 and the larger, material-dependent variation of $l_0$ by 1-2 decades [partly caused by uncertainties from the extrapolation $T \to \infty$, see Fig. 3(a)], Eq. (7) leads to a much smaller variance of $y_m^{id}$ than the observed, huge nine-decades variation of $y_m$ (inset of Fig. 1).

Thus, the latter is mostly due to cooperativity (or any alternative mechanism giving rise to fragility). Using actual values for the parameters in Eq. (7), i.e., $T_g/T_m = 2/3$, $l_g = 2$ or $12$, and $l_0 = -14$ or $-4$ (for $\tau$ and $\eta$, respectively), leads to $\log_{10} \tau_m^{id} = -3.33$ and $\log_{10} \eta_m^{id} = 6.67$, in agreement with the results derived above from Fig. 3.

## IV. EXPERIMENTAL EXAMPLES

Figure 4 shows two experimental examples using $\tau(T)$ data for the molecular glass former glycerol [23] [Fig. 4(a)] and the polymer poly(ethylene oxide) [PEO; Fig. 4(b)] [39,41] (open symbols). The solid lines are fits with the VFT formula, Eq. (1), and the dashed lines show $\tau(T)$ without cooperativity. The evaluation is analogous to Fig. 3(a), however, using experimental values for $T_m$ and $m$ [22,23,39,40] to calculate $T_m^{id}$ via Eq. (4). Glycerol is an intermediate and PEO a fragile liquid and their $\tau_m$ values [23,41] (spheres in Fig. 4) differ by about two decades. In contrast, their $\log_{10} \tau_m^{id}$ values of -3.4 and -4.0, deduced from Fig. 4, vary much less and are compatible with -3.33 derived above. The deviations from this calculated value are due to the differences of the parameters $T_g/T_m = 2/3$, $l_g = 2$, $l_0 = -14$, assumed in its derivation, from the experimental ones. This is especially relevant for $l_0$ which is -14.4 for glycerol and -13.4 for PEO. For both materials, $T_m$ and $T_m^{id}$ differ considerably, but this difference is more pronounced for the fragile PEO, revealing the strong influence of cooperativity on the melting temperature. In the SM (Fig. S3) [1], analogous results are documented for another molecular and polymeric glass former. Moreover, in Fig. S4 [1] experimental examples are provided for viscosity data of three network glass formers and a metallic system. Again, strongly varying values of $\eta_m$ for the different materials (by 6.6 decades) are modified into much less varying values of $\eta_m^{id}$ if one corrects for the influence of fragility. As discussed above, the remaining variation of about 2.8 decades is due to the fact that the VFT does not always lead to perfect fits, the 2/3 factor reveals some scatter [Fig. 2(b)], and the values of $l_g$ and $l_0$ differ somewhat for different materials.

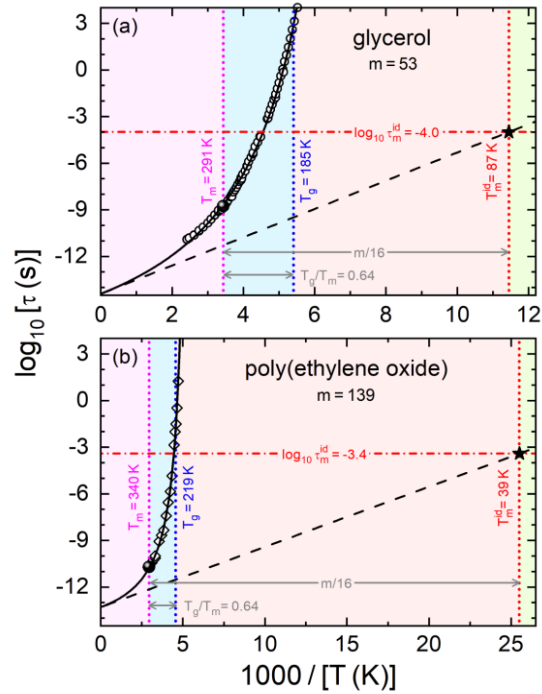

FIG. 4. Symbols: Arrhenius plots of the temperature-dependent $\tau$ of glycerol [23] (a) and PEO [39,41] (b). Solid lines: fits with Eq. (1). $m$ was derived from $D$ [2]. The two leftmost dotted lines indicate experimental values of $T_m$ and $T_g$ [22,23,40] and the lower double arrows visualize $T_g/T_m$. The dashed lines show $\tau(T)$ without fragility [Eq. (5)]. The rightmost dotted lines indicate $T_m^{id}$, determined from the experimental $T_m$ using Eq. (4) (the factor $m/16$ is indicated by the upper double arrows). The dash-dotted lines and stars represent the resulting $\tau_m^{id}$. The spheres show $\tau_m$.

## V. SUMMARY AND CONCLUSIONS

In conclusion, the experimentally founded existence of an approximately fixed $T_g/T_m$ ratio (Fig. 2) [11,22,42,49,65] implies a fundamental relaxation time and viscosity for fragility-free melting. Notably, this statement is completely independent of the somewhat arbitrary definition of $T_g$, which essentially depends on the average lifetime (or patience) of the observing species [77]. Using the usual definitions $\tau(T_g) = 100$ s and $\eta(T_g) = 10^{12}$ Pa s and making typical assumptions for the involved parameters, we obtain the universal values $\tau_m^{id} = 10^{-3.33}$ s and $\eta_m^{id} = 10^{6.67}$ Pa s. This "idealized" melting is only governed by the Lindemann criterion and, for fragile liquids, would lead to much smaller melting temperatures than real-world melting. However, the latter is also strongly affected by the mechanism causing fragility in the liquid phase, e.g., the cooperativity of particle motion, which considerably enhances $T_m$. The experimentally observed strong variation of $\eta$ and $\tau$ at $T_m$ and the difference of real-world and Lindemann melting can be completely ascribed to this fragility effect and, on a microscopic level, thus are explained by the material-dependent degree of cooperativity of molecular motions. Notably, it is fortunate that the melts of the widely applied silicate glasses are rather strong liquids, as otherwise the temperatures needed for their production from the crystalline raw materials would be much higher. In this respect, strong materials should generally be preferable for applications. As there is no reason to assume that the concepts of cooperativity and fragility only apply to liquids that can be easily supercooled [42], we want to point out that the considerations of the present work are valid for all liquids, irrespective of their glass-forming abilities.

In any case, for a deeper understanding of the present findings, a theoretical explanation for one of the closely interrelated values $\tau_m^{id} \approx 10^{-3.33}$ s, $\eta_m^{id} \approx 10^{6.67}$ Pa s, $T_g/T_m \approx 2/3$ or $E/T_m^{id} \approx 24.6$ is desirable. The latter one seems the most fundamental as it connects two, at first glance unrelated energy scales: the average barrier $E$, relevant for fragility-free, relaxational particle motions within the liquid, and the energy $k_B T_m^{id}$ which (within the Lindemann picture) is characteristic for vibrational motions that are sufficiently vigorous to trigger idealized melting without fragility.

## ACKNOWLEDGMENTS

We thank Alessio Zaccone for stimulating discussions.

# Supplementary Material

for

# The 2/3 Rule of Glass Physics Implies Universalities in Crystal Melting

Peter Lunkenheimer[1,*], Konrad Samwer[2], and Alois Loidl[1]

[1] *Experimental Physics V, Center for Electronic Correlations and Magnetism, University of Augsburg, 86135 Augsburg, Germany*
[2] *1. Physikalisches Institut, University of Göttingen, 37077 Göttingen, Germany*

* Contact author: peter.lunkenheimer@physik.uni-augsburg.de

TABLE S1. Material parameters. Fragility indices $m$, melting temperatures $T_m$, relaxation times $\tau_m$ at $T_m$, and viscosities $\eta_m$ at $T_m$ of various glass formers as used in the inset of Fig. 1.

|  | $m$ | $T_m$ (K) | $\tau_m$ (s) | $\eta_m$ (Pa·s) |
|---|---|---|---|---|
| $SiO_2$ | 20 [1] | 2003 [2] |  | $5.0 \times 10^6$ [3] |
| Albite | 22 [4] | 1373 [5] |  | $1.1 \times 10^7$ [6] |
| $BeF_2$ | 24 [7] | 821 [2] |  | $5.6 \times 10^5$ [8] |
| $ZnCl_2$ | 30 [1] | 590 [2] |  | 5.0 [3] |
| $B_2O_3$ | 32 [1] | 793 [2] |  | 2340 [9] |
| $As_2Se_3$ | 36 [7] | 645 [10] |  | 9 [11] |
| $Zr_{41.2}Ti_{13.8}Cu_{12.5}Ni_{10}Be_{22.5}$ | 39 [12] | 937 [13] |  | 34 [14] |
| $Zr_{46.75}Ti_{8.25}Cu_{7.5}Ni_{10}Be_{27.5}$ | 43 [13] | 1050 [13] |  | 17 [15] |
| $Na_2Si_2O_5$ | 45 [1] | 1147 [16] |  | 900 [17] |
| $Zr_{11}Cu_{47}Ti_{34}Ni_8$ | 47 [18] | 1160 [19] |  | 0.4 [20] |
| Glycerol | 53 [1] | 291 [21] | $1.7 \times 10^{-9}$ [22] | 1.7 [23] |
| Diopside | 53 [4] | 1670 [24] |  | 2.8 [6] |
| Salol | 73 [1] | 315 [25] | $2.1 \times 10^{-10}$ [22] |  |
| Polybutadiene | 85 [26] | 285 [27] | $1 \times 10^{-9}$ [28] |  |
| Sorbitol | 93 [1] | 388 [29] | $1.1 \times 10^{-10}$ [22] | 0.44 [30] |
| Bmim Cl | 97 [31] | 330 [32] | $6.5 \times 10^{-10}$ [31] | 0.80 [33] |
| Xylitol | 103 [22][a] | 366 [34] | $2.0 \times 10^{-10}$ [22] |  |
| Propylene carbonate | 104 [1] | 224 [35] | $5.9 \times 10^{-10}$ [22] | $4.7 \times 10^{-2}$ [23] |
| Benzophenone | 125 [36] | 321 [25] | $1.1 \times 10^{-10}$ [36] | $5.9 \times 10^{-3}$ [37] |
| Poly(ethylene oxide) | 139 [38] | 340 [39] | $2.0 \times 10^{-11}$ [40] |  |

[a]Calculated from the strength parameter $D$ given in Ref. [22].

## Derivation of Eq. (3) of main text (fragility dependence of $\eta$ and $\tau$ at $T_m$)

The derivation is done here for $\tau$, for $\eta$ the calculations are analogous (see end of section).

Take the VFT formula, Eq. (1) of the main text:

$$y = y_0 \exp\left(\frac{DT_{VF}}{T - T_{VF}}\right)$$

Logarithmic version for $\tau$ using $\tau_0 = 10^{-14}$ s:

$$\log_{10} \tau = -14 + \frac{1}{\ln 10} \frac{DT_{VF}}{T - T_{VF}}$$

For $T = T_g \to \tau(T_g) = 100$ s:

$$16 = \frac{1}{\ln 10} \frac{DT_{VF}}{T_g - T_{VF}}$$

$$T_g = T_{VF}\left(1 + \frac{D}{16 \ln 10}\right) \tag{S1}$$

Logarithmic VFT formula at $T_m = 3/2\ T_g$ [Eq. (2)] using $\tau_m := \tau(T_m)$:

$$\log_{10} \tau_m = -14 + \frac{1}{\ln 10} \frac{DT_{VF}}{3/2\ T_g - T_{VF}}$$

With $T_g$ from Eq. (S1) $\to$

$$\log_{10} \tau_m = -14 + \frac{1}{\ln 10} \frac{DT_{VF}}{3/2\ T_{VF}\left(1 + \frac{D}{16 \ln 10}\right) - T_{VF}}$$

$$= -14 + \frac{1}{\ln 10} \frac{D}{\frac{3/2\ D}{16 \ln 10} + 0.5} = -14 + \frac{1}{\ln 10} \frac{D}{\frac{3/2\ D + 0.5 \cdot 16 \ln 10}{16 \ln 10}} = -14 + \frac{1}{\ln 10} \frac{D \cdot 16 \ln 10}{3/2\ D + 8 \ln 10}$$

$$\log_{10} \tau_m = -14 + \frac{16\ D}{3/2\ D + 8 \ln 10}$$

Using $D = 16^2 \ln 10 / (m-16)$ derived from Eq. (4) in Ref. [1] $\to$

$$\log_{10} \tau_m = -14 + \frac{16^3 \ln 10}{(m - 16)\left(3/2\ 16^2 \frac{\ln 10}{m - 16} + 8 \ln 10\right)} = -14 + \frac{16^2}{3/2\ 16 + m/2 - 8}$$

$$\log_{10} \tau_m = \frac{16^2}{16 + m/2} - 14$$

Considering that $m_{min} = 16$, this is Eq. (3) of the main text for $y = \tau$.

For the viscosity, analogous calculations (assuming $\eta(T_g) = 10^{12}$ Pa·s and $\eta_0 = 10^{-4}$ Pa·s) lead to:

$$\log_{10} \eta_m = \frac{16^2}{16 + m/2} - 4$$

# Derivation of Eq. (7) of main text (calculation of $y_\mathrm{m}^\mathrm{id}$)

## 1. $y(T)$ with and without cooperativity

With cooperativity: VFT behavior (solid lines in Fig. 2)

$$y = y_0 \exp\left(\frac{DT_\mathrm{VF}}{T-T_\mathrm{VF}}\right) \quad \text{(Eq. (1) in main text)} \qquad (S2)$$

With cooperativity "switched off": dashed Arrhenius-lines in Fig. 2 (see main text):

$$y = y_0 \exp\left(\frac{DT_\mathrm{VF}}{T}\right) \quad \text{(Eq. (5) in main text)} \qquad (S3)$$

($y$ stands for $\eta$ or $\tau$)

## 2. Two simplifying definitions (see Fig. S1):

$l_0 := \log_{10} y_0$

   For $y = \tau$: usually $l_0 = -14$
   For $y = \eta$: $l_0 = -4$ or $-5$
   ($l_0 = -4$ retains 16 as minimum $m$, see remark in main text)

$l_\mathrm{g} := \log_{10}[y(T_\mathrm{g})]$

   For $y = \tau$: usually $l_\mathrm{g} = 2$
   For $y = \eta$: usually $l_\mathrm{g} = 12$

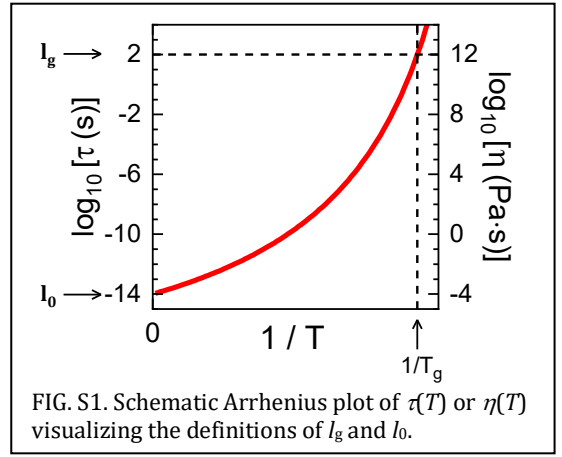

FIG. S1. Schematic Arrhenius plot of $\tau(T)$ or $\eta(T)$ visualizing the definitions of $l_\mathrm{g}$ and $l_0$.

## 3. Ideal melting temperature

The experimental data on expansion coefficient versus $T_\mathrm{m}$ in Ref. [41] suggest:

$$T_\mathrm{m} = \frac{m}{16} T_\mathrm{m}^\mathrm{id} \quad \text{(Eq. (4) in main text)}$$

Use more universal expression:

$$T_\mathrm{m} = \frac{m}{l_\mathrm{g}-l_0} T_\mathrm{m}^\mathrm{id} \qquad (S4)$$

$T_\mathrm{m}^\mathrm{id}$: hypothetical melting temperature with cooperativity "switched off" (see main text).
$y_\mathrm{m}^\mathrm{id} := y(T_\mathrm{m}^\mathrm{id})$ (as defined in main text)

## 4. Relation of $D$ and $m$

Ref. [1] (assuming $l_\mathrm{g} = 2$, $l_0 = -14$):   $m = 16 + 16^2 \ln 10 / D$

More universal version:   $m = l_\mathrm{g} - l_0 + (l_\mathrm{g} - l_0)^2 \ln 10 / D$

$\rightarrow$ 
$$D = \frac{(l_\mathrm{g}-l_0)^2 \ln 10}{m - l_\mathrm{g} + l_0} \qquad (S5)$$

## 5. $T = T_g$

Logarithmic version of Eq. (S2): 
$$\log_{10} y = l_0 + \frac{1}{\ln 10} \frac{DT_{VF}}{T - T_{VF}}$$

$T = T_g \rightarrow$ 
$$l_g = l_0 + \frac{1}{\ln 10} \frac{DT_{VF}}{T_g - T_{VF}}$$

$$(l_g - l_0) \ln 10 = \frac{DT_{VF}}{T_g - T_{VF}}$$

(S5) $\rightarrow$ 
$$(l_g - l_0) \ln 10 = \frac{(l_g - l_0)^2 \ln 10 \, T_{VF}}{(m - l_g + l_0)(T_g - T_{VF})}$$

$$1 = \frac{(l_g - l_0) T_{VF}}{(m - l_g + l_0)(T_g - T_{VF})}$$

$$T_g - T_{VF} = \frac{(l_g - l_0) T_{VF}}{m - l_g + l_0}$$

$$T_g = T_{VF}\left(1 + \frac{(l_g - l_0) T_{VF}}{m - l_g + l_0}\right) = T_{VF}\left(\frac{m - l_g + l_0 + l_g - l_0}{m - l_g + l_0}\right) = T_{VF}\left(\frac{m}{m - l_g + l_0}\right)$$

$\rightarrow$ 
$$T_{VF} = T_g \left(\frac{m - l_g + l_0}{m}\right) \tag{S6}$$

## 6. $T = T_m^{id}$

Logarithmic version of Eq. (S3): 
$$\log_{10} y = l_0 + \frac{1}{\ln 10} \frac{DT_{VF}}{T}$$

$T = T_m^{id} \rightarrow$ 
$$\log_{10} y_m^{id} = l_0 + \frac{1}{\ln 10} \frac{DT_{VF}}{T_m^{id}}$$

(S4) $\rightarrow$ 
$$\log_{10} y_m^{id} - l_0 = \frac{1}{\ln 10} \frac{DT_{VF}}{T_m(l_g - l_0)/m}$$

(S5) $\rightarrow$ 
$$\log_{10} y_m^{id} - l_0 = \frac{1}{\ln 10} \frac{(l_g - l_0)^2 \ln 10 \, T_{VF} \, m}{(m - l_g + l_0) T_m (l_g - l_0)} = \frac{(l_g - l_0) T_{VF} \, m}{(m - l_g + l_0) T_m}$$

(S6) $\rightarrow$ 
$$\log_{10} y_m^{id} - l_0 = \frac{(l_g - l_0) T_g \frac{m - l_g + l_0}{m} m}{(m - l_g + l_0) T_m}$$

## 7. Final results:

$\rightarrow$ 
$$\boxed{\log_{10} y_m^{id} = \frac{T_g}{T_m}(l_g - l_0) + l_0} \tag{S7}$$

(Eq. (7) of the main text)

$\rightarrow$ 
$$\frac{T_g}{T_m} = \frac{\log_{10} y_m^{id} - l_0}{l_g - l_0} \tag{S8}$$

8. Eq. (S7) with actual values for $T_g/T_m$, $l_g$, $l_0$

a) For $y = \tau$:

Assumptions:
1. $T_g/T_m = 2/3$ (experimentally verified for most materials: 2/3 ±0.15; see Fig. S2)
2. $\tau(T_g) = 100$ s ($l_g = 2$) (definition)
3. $\tau_0 = 10^{-14}$ s ($l_0 = -14$) (approximately valid, 1-2 decades variation)

(S7) → $$\log_{10} \tau_m^{id} = {}^2\!/\!_3\, 16 - 14 = -3.33$$

b) For $y = \eta$:

Assumptions:
1. $T_g/T_m = 2/3$
2. $\eta(T_g) = 10^{12}$ Pa·s ($l_g = 12$)
3. $\eta_0 = 10^{-4}$ Pa·s ($l_0 = -4$)

(S7) → $$\log_{10} \eta_m^{id} = {}^2\!/\!_3\, 16 - 4 = 6.67$$

Alternatively:
3. $\eta_0 = 10^{-5}$ Pa·s ($l_0 = -5$):

(S7) → $$\log_{10} \eta_m^{id} = {}^2\!/\!_3\, 17 - 5 = 6.33$$

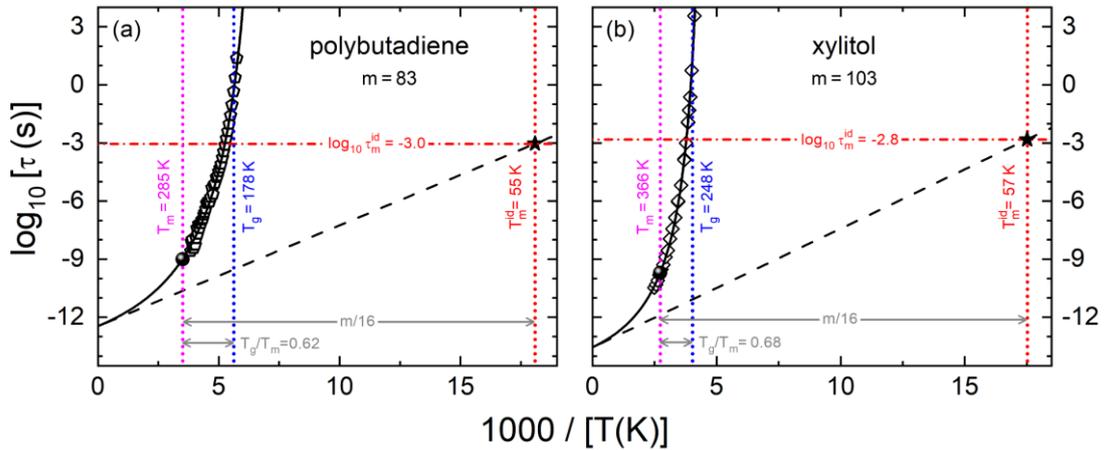

FIG. S2. Further experimental examples of melting-point enhancement by fragility using $\tau(T)$ data. Symbols: Arrhenius plots of the temperature-dependent relaxation times of (a) polybutadiene [28] and (b) xylitol [22]. Experimental values of $T_g$ and $T_m$ are taken from Refs. [28] and [27] (polybutadiene) and from Refs. [22] and [34] (xylitol). For the meaning of the symbols and lines, see caption of Fig. 3.

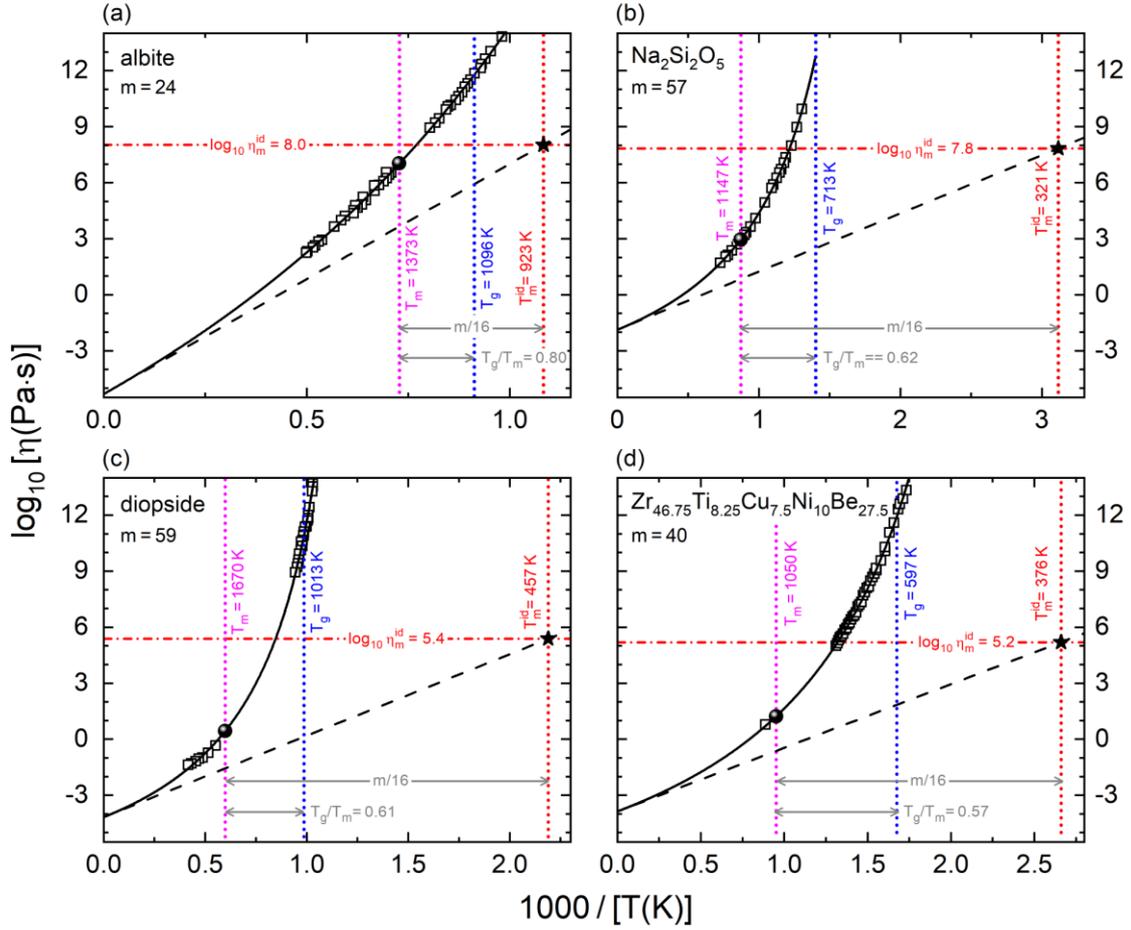

FIG. S3. Further experimental examples of melting-point enhancement by fragility using $\eta(T)$ data. Symbols: Arrhenius plots of the temperature-dependent viscosities of three network glass formers and a metallic system, namely: (a) albite [6], (b) $Na_2Si_2O_5$ [17], (c) diopside [6], and (d) $Zr_{46.75}Ti_{8.25}Cu_{7.5}Ni_{10}Be_{27.5}$ [15]. Experimental values of $T_g$ and $T_m$ are taken from Refs. [6] and [5] (albite), [3] and [16] ($Na_2Si_2O_5$), [42] and [24] (diopside), and from Ref. [13] ($Zr_{46.75}Ti_{8.25}Cu_{7.5}Ni_{10}Be_{27.5}$). For the meaning of the lines, spheres, and stars, see caption and discussion of Fig. 3, where an analogous analysis was performed for $\tau(T)$ data (note that for $\eta$ data, the calculated value of $\log_{10}\eta_m^{id}$ is 6.67).